\documentclass[12pt]{article}
\usepackage{amsfonts}
\usepackage{latexsym}
\usepackage{amsmath}
\usepackage{amssymb}
\usepackage{amssymb}

\newcommand{\newsection}{    
\setcounter{equation}{0}\section}
\def\appendix#1{\addtocounter{section}{1}\setcounter{equation}{0}
\renewcommand{\thesection}{\Alph{section}}
\section*{Appendix \thesection\protect\indent \parbox[t]{11.15cm}{#1}}
\addcontentsline{toc}{section}{Appendix \thesection\ \ \ #1}}

\font\mybb=msbm10 at 11pt

\def\bb#1{\hbox{\mybb#1}}

\def\bR {\bb{R}}
\def\bC {\bb{C}}

\begin{document}
\begin{titlepage}
\begin{center}

\vspace{5.0cm}

\vfill

\begin{center}
   \baselineskip=16pt
  {\Large\bf Small Horizons}
   \vskip 2cm
       Jan B. Gutowski$^1$, Dietmar Klemm$^{2,3}$, Wafic Sabra$^4$ and Peter Sloane$^3$\\
 \vskip .6cm
      \begin{small}
      $^1$\textit{Department of Mathematics, King's College London
      \\ Strand, London WC2R 2LS, United Kingdom}
        \end{small}\\*[.6cm]
      \begin{small}
      $^2$\textit{ Universit\`a di Milano, Dipartimento di Fisica, \\
\hspace*{0.15cm} Via Celoria 16, 20133 Milano, Italy}
        \end{small}\\*[.6cm]
        \begin{small}
      $^3$\textit{INFN, Sezione di Milano, Via Celoria 16, 20133 Milano, Italy }
        \end{small}\\*[.6cm] 
      \begin{small}
      $^4$\textit{Centre for Advanced Mathematical Sciences and
        Physics Department, \\
        American University of Beirut, Lebanon}
        \end{small}      
   \end{center}

\vspace{0.5cm}

\end{center}
{}
\vskip 2.0 cm
\begin{abstract}

\vskip5mm

All near horizon geometries of supersymmetric black holes in a $N=2, D=5$ higher-derivative supergravity theory are
classified. Depending on the choice of near-horizon data we find that either there are no regular horizons,
or horizons exist and the spatial cross-sections of the event horizons are  conformal to a squashed or round $S^3$, $S^1 \times S^2$,
or $T^3$.  If the conformal factor is constant then
the solutions are maximally supersymmetric. If the conformal factor is not constant, we find that it  satisfies a non-linear  vortex equation, 
and the horizon may admit scalar hair.

\end{abstract}
\end{titlepage}


\newsection{Introduction}

In recent years it has become evident that there are exotic black hole
solutions in higher dimensional gravitational theories. The most notable
examples are the five-dimensional black rings \cite{ring1, ring2, ring3,
ring4, ring5, ring6}. These are solutions where the spatial cross-sections
of the event horizon have $S^{1}\times S^{2}$ topology. Moreover, the black
hole uniqueness theorems, originally formulated in four dimensional general
relativity \cite{unique1, unique2, unique3, unique4, unique5, unique6}, do
not generalize straightforwardally to higher dimensions. However, uniqueness
theorems have been formulated for static solutions in higher dimensions in  
\cite{gibbons1, rogatko}, and for solutions with extra rotational Killing
vectors, in \cite{extr3, extr4, extr5}. One method to investigate the
structure of extremal higher dimensional black objects with regular horizons
is to study their near-horizon limit. In such a limit, information about the
asymptotic behaviour of the black hole is removed and only information
concerning the structure of the horizon is retained. If one considers
supersymmetric black holes, then further conditions on the near-horizon
geometry are obtained due to supersymmetry. Supersymmetric near-horizon
geometries for the ungauged five-dimensional minimal supergravity were first
considered in \cite{reallbh}. The case with vector multiplets was considered
in \cite{gutbh}. Later in \cite{adsbh}, the results of \cite{reallbh} were
generalized to the minimal gauged supergravity with negative cosmological
constant. In this case, one obtains weaker conditions and as such a complete
classification of the near-horizon geometries was not possible. However, new
solutions were found which were subsequently generalized in \cite{pope}.
Also supersymmetric near-horizon geometries, with two commuting rotational
Killing vectors, in the theory with a negative cosmological constant, were
considered in \cite{kunduri1}. The near horizon analysis was also performed
for ten-dimensional heterotic supergravity in \cite{heterotic}. We note that
the near-horizon geometries of the so called five-dimensional de-Sitter
supergravity theory coupled to vector multiplet was performed recently in 
\cite{cosmorings}.

In the present work we shall investigate the near-horizon geometries of supersymmetric extremal 
black hole solutions in higher derivative $N=2,$ $D=5$ supergravity, coupled
to a number of abelian vector multiplets \cite{hanaki}. The
higher-derivative theory has, in addition to the spacetime metric, real
scalars $X^{I}$, abelian 2-form field strengths $F^{I}$, and two auxiliary
fields consisting of an auxiliary 2-form $H$ and a real auxiliary scalar $D$.
The solutions found are either the maximally supersymmetric near horizon
solutions found in \cite{reallbh, katmadas}, or solutions for which the spatial
cross-sections of the event horizon are conformal to a squashed or round $S^{3},$ 
$S^{1}\times S^{2}$ or $T^{3}.$ The function defining the conformal factor
satisfies a non linear partial differential equation.

This work is organised as follows. In section two, necessary and sufficient
conditions for the existence of \ a supersymmetric near-horizon geometry
associated with the event horizon of a supersymmetric extremal black hole in
our theory are examined. In sections three and four the local conditions
satisfied by our geometries are obtained via the analysis of the gravitino,
gaugino and auxiliary Killing spinor equations. In section five we perform
the global analysis by demanding that the spatial cross-section of the event
horizon $\mathcal{S}$ is compact without boundary. It is demonstrated that 
$\mathcal{S}$ must be conformal to one of these spaces: squashed $S^3$,
round $S^3$, $S^{1}\times S^{2}$ and $T^{3}.$ In section six, we consider
the auxiliary $D$-field equation. It turns out that if this equation is
satisfied, together with the conditions obtained from the Killing spinor
equations, then all the remaining equations of motion are satisfied. 
The $D$-equation of motion implies that either there are no solutions, or the solutions reduce to
those found in \cite{reallbh, gutbh, katmadas}, or the conformal factor satisfies a vortex-like
nonlinear partial differential equation. In section 7, we introduce local co-ordinates and 
list all of the solutions. We conclude in section 8.

\newsection{Supersymmetry and Near-Horizon Geometries}

We shall examine the necessary and sufficient conditions for there to be
a supersymmetric near-horizon geometry associated with the event horizon
of a supersymmetric extremal black hole in higher derivative ungauged
$N=2, D=5$ supergravity coupled to an arbitrary number of abelian vector
multiplets. After taking the near-horizon limit, the metric on the near-horizon geometry
is \cite{reallbh, gnull1, gnull2}
\begin{eqnarray}
\label{met}
ds^2 =2 du (dr+rh-{1 \over 2}r^2 \Delta du) + ds^2_{{\cal{S}}} \ .
\end{eqnarray}
Here ${\partial \over \partial u}$ is a Killing vector; it is assumed that the event horizon
is a Killing horizon of ${\partial \over \partial u}$. This has been shown to hold for a large class of 2-derivative 
supergravity theories coupled to Maxwell fields and scalars \cite{symm3}, modulo certain technical assumptions,
however it has not been proven for the higher derivative theory we consider here. 

The horizon is located at $r=0$,
and ${\cal{S}}$ denotes the spatial cross-sections of the event horizon, which is taken to be compact
and without boundary. The metric $ds^2_{\cal{S}}$ does not depend on $u$ or $r$,
$\Delta$ and $h$ are a scalar and 1-form on ${\cal{S}}$ respectively, which also do not
depend on $u$ or $r$. We remark that the near-horizon limit corresponds to setting
\begin{eqnarray}
r = \lambda r', \quad u = \lambda^{-1} u'
\end{eqnarray}
and then taking the limit $\lambda  \rightarrow 0$ and dropping the primes.

We shall mostly use the conventions of
\cite{larsen}, however we denote the scalars $M^I$ as $X^I$,
and rescale the auxiliary 2-form field $v$ as $v= {3 \over 4} H$
in order to simplify some coefficients. We also work in a mostly plus signature ($-$,$+$,$+$,$+$,$+$).
With these modified conventions, the gravitino, gaugino and auxiliary Killing spinor equations (KSEs) are
\begin{eqnarray}
\label{grav}
\nabla_\mu \epsilon -{i \over 8} \Gamma_\mu H_{\nu_1 \nu_2} \Gamma^{\nu_1 \nu_2}
\epsilon +{3 i \over 4} H_\mu{}^\nu \Gamma_\nu \epsilon =0
\end{eqnarray}
and
\begin{eqnarray}
\label{gaug}
\bigg(\big(F^I+X^I H\big)_{\nu_1 \nu_2} \Gamma^{\nu_1 \nu_2}
+2i \Gamma^\nu \nabla_\nu X^I  \bigg) \epsilon =0
\end{eqnarray}
and
\begin{eqnarray}
\label{aux}
\bigg( D -{3 \over 2} H_{\nu_1 \nu_2} H^{\nu_1 \nu_2}
-{i \over 2} dH_{\nu_1 \nu_2 \nu_3} \Gamma^{\nu_1 \nu_2 \nu_3}
+{3 i \over 2} \star \big( d \star H + H \wedge H \big)_\nu \Gamma^\nu \bigg) \epsilon =0
\end{eqnarray}
where $\epsilon$ is a Dirac Killing spinor whose structure will be investigated in greater detail later,
and
\begin{eqnarray}
\nabla_\mu \epsilon = \big(\partial_\mu +{1 \over 4} \omega_{\mu, \nu_1 \nu_2} \Gamma^{\nu_1 \nu_2} \big) \epsilon
\end{eqnarray}
is the supercovariant derivative, where $\omega$ is the spin connection.
It will be convenient to work with a light-cone basis $\{ {{\bf{e}}}^+, {{\bf{e}}}^-, {{\bf{e}}}^i \}$ for $i=1,2,3$
such that ${{\bf{e}}}^i$ is a ($u$, $r$-independent) basis for ${\cal{S}}$ and
\begin{eqnarray}
\label{basis}
{{\bf{e}}}^+ = du, \qquad {{\bf{e}}}^- = dr+ rh -{1 \over 2} r^2 \Delta du
\end{eqnarray}
and
\begin{eqnarray}
\label{frame}
ds^2=2 {{\bf{e}}}^+ {{\bf{e}}}^- + \delta_{ij} {{\bf{e}}}^i {{\bf{e}}}^j \ .
\end{eqnarray}
The non-vanishing components of the spin  connection associated with this basis are listed in
Appendix A.
In addition to the metric ({\ref{met}}) being regular in the near-horizon limit, we shall furthermore
assume that all of the other bosonic fields are also regular in this limit, including the auxiliary fields $H, D$.
In terms of the scalars, this means that after taking the near-horizon limit,
$X^I$ and $D$ are smooth functions  on ${\cal{S}}$ which are independent of $u$ and $r$.
Furthermore, the 2-forms $H$ and $F^I$ can be written as
\begin{eqnarray}
H = \Phi {{\bf{e}}}^+ \wedge {{\bf{e}}}^- + r {{\bf{e}}}^+ \wedge {\cal{B}} + {\tilde{H}}
\end{eqnarray}
and
\begin{eqnarray}
F^I = \Phi^I {{\bf{e}}}^+ \wedge {{\bf{e}}}^- + r {{\bf{e}}}^+ \wedge {\cal{B}}^I + {\tilde{F}}^I
\end{eqnarray}
where $\Phi, \Phi^I$ are smooth $u, r$-independent scalars on ${\cal{S}}$;
${\cal{B}}, {\cal{B}}^I$ are smooth $u, r$-independent 1-forms on ${\cal{S}}$;
and ${\tilde{H}}, {\tilde{F}}^I$ are smooth $u, r$-independent 2-forms on ${\cal{S}}$.

We shall also find it convenient to decompose spinors into positive and negative chirality parts
\begin{eqnarray}
\epsilon = \epsilon_++ \epsilon_-, \qquad \Gamma_\pm  \epsilon_\pm =0,
\qquad \Gamma_\pm \epsilon_\pm = \pm \epsilon_\pm
\end{eqnarray}
and note the useful identities
\begin{eqnarray}
\Gamma_{ij} \epsilon_\pm = \mp i \epsilon_{ij}{}^k \Gamma_k \epsilon_\pm,
\quad \Gamma_{ijk} \epsilon_\pm = \mp i \epsilon_{ijk} \epsilon_\pm
\end{eqnarray}
where $\epsilon_{ijk}$ denotes the volume form of ${\cal{S}}$.
Various spinorial geometry conventions are listed in Appendix B

\newsection{Analysis of Gravitino KSE}

To begin with, we analyse the gravitino KSE ({\ref{grav}}). As all of the dependence on the $u$, $r$
components in the bosonic fields is known explicitly, we begin by solving the $+$ and the $-$ components.
In particular, from the $-$ component one finds that
\begin{eqnarray}
\label{solv1}
\epsilon_+ = \phi_+, \qquad  \epsilon_- = \phi_- + r \Gamma_- \big({1 \over 4}(h+\star_3 {\tilde{H}})_i \Gamma^i
+{i \over 2} \Phi \big) \phi_+
\end{eqnarray}
where $\phi_\pm$ do not depend on $r$, and $\star_3$ denotes the Hodge dual on ${\cal{S}}$.
The $+$ component of the KSE implies that
\begin{eqnarray}
\label{plus1}
\partial_u \epsilon_+  + \big({1 \over 2} r \Delta
+{i \over 4} r (\star_3 dh)_i \Gamma^i +{i \over 4} r {\cal{B}}_i \Gamma^i \big) \epsilon_+
+ \Gamma_+ \big(-{1 \over 4}(h-\star_3 {\tilde{H}})_i \Gamma^i +{i \over 2} \Phi \big) \epsilon_-=0
\end{eqnarray}
and
\begin{eqnarray}
\label{plus2}
\partial_u \epsilon_-  + \big(-{1 \over 2} r \Delta -{i \over 4}r (\star_3 dh)_i \Gamma^i
+{3i \over 4} r {\cal{B}}_i \Gamma^i \big) \epsilon_-
\nonumber \\
+r^2 \Gamma_- \bigg( {1 \over 4} (\Delta h_i - \partial_i \Delta) \Gamma^i
+{1 \over 2} \Delta \big({1 \over 4} (h+\star_3 {\tilde{H}})_i \Gamma^i +{i \over 2}\Phi \big) \bigg) \epsilon_+=0 \ .
\end{eqnarray}
Note that ({\ref{plus1}}) and ({\ref{plus2}}) imply that
\begin{eqnarray}
\label{solv2}
\phi_- = \eta_- , \qquad \phi_+ = \eta_+ + u \Gamma_+  \big({1 \over 4}(h-\star_3 {\tilde{H}})_i \Gamma^i
-{i \over 2} \Phi \big) \eta_-
\end{eqnarray}
where $\eta_\pm$ do not depend on $u$ and $r$, and $\eta_\pm$ must also satisfy a number
of algebraic conditions.

Before considering these algebraic conditions in further detail, it is useful to compute
the 1-form spinor bilinear
\begin{eqnarray}
Z_\mu = -{1 \over 2} B(\epsilon, \Gamma_\mu \epsilon)
\end{eqnarray}
where $B$ is the $Spin(4,1)$ invariant inner product defined in ({\ref{prod}}).
It is known that this 1-form is dual to a Killing vector, which is a symmetry of
the full solution \cite{class1}.
We shall require that this 1-form spinor bilinear be proportional to
\begin{eqnarray}
V = -{1 \over 2} r^2 \Delta {{\bf{e}}}^+ + {{\bf{e}}}^-
\end{eqnarray}
where $V$ is the 1-form dual to the Killing vector ${\partial \over \partial u}$.
We remark that this condition is not a priori necessary. In particular,
it need not hold in the case for which the near-horizon geometry is supersymmetric,
but the bulk black hole solution is not. Such solutions are known to exist in the
2-derivative theory \cite{nonbps1, nonbps2}, and may also exist in the higher derivative theory as well.
However, in this work we shall assume that both the bulk and the near-horizon geometry
are supersymmetric, and therefore take $Z$ to be proportional to $V$.

Recalling that at $r=u=0$, $\epsilon=\eta_+ + \eta_-$ as a consequence of
({\ref{solv1}}) and ({\ref{solv2}}), it is straightforward to show that requiring that $Z_+=0$ at $r=u=0$
implies that
\begin{eqnarray}
\eta_-=0 \ .
\end{eqnarray}
Furthermore, $\eta_+$ can be simplified further by making use of a $r, u$-independent $SU(2)$ gauge transformation to
write
\begin{eqnarray}
\eta_+ = \alpha (1-e_1)
\end{eqnarray}
for some $r, u$ independent function $\alpha$, $\alpha \in \bR$.
It follows that at $r=0$, $\epsilon = \eta_+ = \alpha (1-e_1)$.
Also note that at $r=0$,
\begin{eqnarray}
Z_- = \sqrt{2} \alpha^2
\end{eqnarray}
and so comparing with $V_-$ we require that $\alpha^2$ be constant. Without loss of generality set
$\alpha=1$, so
\begin{eqnarray}
\label{simp1}
\eta_+ = 1-e_1 \ .
\end{eqnarray}
Next, on imposing the conditions $Z_i=0$, one finds that
\begin{eqnarray}
\label{cx1}
{\tilde{H}} = - \star_3 h \ .
\end{eqnarray}
The Killing spinor therefore simplifies further, and one finds that
\begin{eqnarray}
\label{simp2}
\epsilon_+ = \eta_+, \qquad \epsilon_- = {i \over 2} r \Phi \Gamma_- \eta_+
\end{eqnarray}
where $\eta_+$ is given by ({\ref{simp1}}).
Finally, we compute the ratio
\begin{eqnarray}
{Z_+ \over Z_-}= - {r^2 \over 2}\Phi^2 \ .
\end{eqnarray}
On requiring that this be equal to ${V_+ \over V_-}$ one finds that
\begin{eqnarray}
\label{cx2}
\Delta = \Phi^2 \ .
\end{eqnarray}
On substituting the Killing spinor ({\ref{simp2}}) back into ({\ref{plus1}}) and ({\ref{plus2}})
and making use of the conditions ({\ref{cx1}}) and ({\ref{cx2}}) one finds two additional conditions
\begin{eqnarray}
\label{cx3}
{\cal{B}} = - \star_3 dh -2 \Phi h
\end{eqnarray}
and
\begin{eqnarray}
\label{cx4}
d \Delta +2 \Delta h +2 \Phi \star_3 dh =0 \ .
\end{eqnarray}
It remains to evaluate the components of ({\ref{grav}}) along the directions of ${\cal{S}}$, with the spinor
$\epsilon$ given by ({\ref{simp2}}).
One finds the following conditions
\begin{eqnarray}
\label{cx5}
{\hat{\nabla}}_i \eta_+ + \big({i \over 4} \Phi \Gamma_i -{i \over 2} (\star_3 h)_i{}^j \Gamma_j \big) \eta_+=0
\end{eqnarray}
where ${\hat{\nabla}}$ is the supercovariant derivative of ${\cal{S}}$, and
\begin{eqnarray}
\label{cx6}
d \Phi + \Phi h + \star_3 dh =0 \ .
\end{eqnarray}
Observe that ({\ref{cx6}}) together with ({\ref{cx2}}) imply ({\ref{cx4}}).
Furthermore, observe that the integrability condition of ({\ref{cx5}}) implies that the
Ricci tensor of ${\cal{S}}$ is
\begin{eqnarray}
\label{ricci1}
{\hat{R}}_{ij} = \big({1 \over 2} \Phi^2 + h^2 - {\hat{\nabla}}^n h_n \big) \delta_{ij} - {\hat{\nabla}}_{(i} h_{j)} - h_i h_j \ .
\end{eqnarray}

To summarize, the gravitino KSE implies that one can take the Killing spinor
$\epsilon$ as in ({\ref{simp2}}), with $\eta_+$ in ({\ref{simp1}}), and in addition,
({\ref{cx1}}), ({\ref{cx2}}), ({\ref{cx3}}), ({\ref{cx5}}) and ({\ref{cx6}}) are obtained,
which in turn imply that the Ricci tensor of ${\cal{S}}$ is given by ({\ref{ricci1}}).
This exhausts the content of ({\ref{grav}}).

\newsection{Analysis of Gaugino and Auxiliary KSE}

Next, we examine the gaugino KSE ({\ref{gaug}}).
On using the conditions obtained in the previous section, one finds that the
positive chirality part of the gaugino KSE implies
\begin{eqnarray}
\Phi^I = - \Phi X^I, \qquad {\tilde{F}}^I = X^I \star_3 h + \star_3 d X^I
\end{eqnarray}
and the negative chirality part of the gaugino KSE implies
\begin{eqnarray}
{\cal{B}}^I  = - X^I {\cal{B}} - \Phi dX^I
\end{eqnarray}
and hence
\begin{eqnarray}
\label{cx7}
F^I = -du \wedge d \big(r \Phi X^I) + X^I \star_3 h + \star_3 d X^I \ .
\end{eqnarray}
Note that the Bianchi identity $dF^I=0$ implies that
\begin{eqnarray}
\label{bian}
{\hat{\nabla}}^2 X^I + h^i {\hat{\nabla}}_i X^I + X^I {\hat{\nabla}}^i h_i =0 \ .
\end{eqnarray}

Next, consider the auxiliary KSE ({\ref{aux}}). To evaluate the condition obtained from this equation,
note that
\begin{eqnarray}
H = du \wedge d(r \Phi) - \star_3 h
\end{eqnarray}
so
\begin{eqnarray}
dH = -d \star_3 h
\end{eqnarray}
and also note that
\begin{eqnarray}
\star \big( d \star H + H \wedge H) = -2r \big( {\hat{\nabla}}^i {\hat{\nabla}}_i \Phi + h^i {\hat{\nabla}}_i \Phi \big) {{\bf{e}}}^+ \ .
\end{eqnarray}
On substituting these conditions into ({\ref{aux}}), one finds that the auxiliary KSE is equivalent to
\begin{eqnarray}
\label{cx8}
D = 3 h^2 -3 \Phi^2-3 {\hat{\nabla}}^i h_i \ .
\end{eqnarray}

\newsection{Global Analysis}

Having extracted all of the local conditions from the KSE, we proceed to obtain additional conditions
by making use of the fact that ${\cal{S}}$ is compact without boundary. In particular, the condition
on the Ricci tensor ({\ref{ricci1}}) implies that ${\cal{S}}$ admits a Gauduchon-Tod structure \cite{tod1, gaud}. There exists
a regular, positive function $\Omega$, such that on making a conformal
re-scaling and setting
\begin{eqnarray}
ds^2_{\tilde{{\cal{S}}}} = \Omega^2 ds^2_{{\cal{S}}}, \qquad h' = h+ \Omega^{-1} d \Omega
\end{eqnarray}
one can choose $\Omega$ such that
\begin{eqnarray}
\label{newgauge}
{\tilde{\nabla}}^i h'_i =0
\end{eqnarray}
where ${\tilde{\nabla}}$ denotes the Levi-Civita connection on ${\tilde{{\cal{S}}}}$ equipped with the conformally rescaled metric
$ds^2_{\tilde{{\cal{S}}}}$. Note that ({\ref{newgauge}}) can be rewritten as
\begin{eqnarray}
\label{conf1}
{\hat{\nabla}}^2 \Omega + h^i {\hat{\nabla}}_i \Omega+ \Omega {\hat{\nabla}}^i h_i =0 \ .
\end{eqnarray}
Furthermore, observe that ({\ref{cx6}}) implies that
\begin{eqnarray}
\label{conf2}
{\hat{\nabla}}^2 \Phi + h^i {\hat{\nabla}}_i \Phi+ \Phi {\hat{\nabla}}^i h_i =0 \ .
\end{eqnarray}
Then ({\ref{conf1}}) and ({\ref{conf2}}) imply
\begin{eqnarray}
{\tilde{\nabla}}^2 (\Phi \Omega^{-1} \big) + (h')^i {\tilde{\nabla}}_i (\Phi \Omega^{-1})=0
\end{eqnarray}
where in the above expression, the frame indices are taken w.r.t the conformally rescaled frame.
Compactness of ${\tilde{{\cal{S}}}}$, then implies that
\begin{eqnarray}
\Phi \Omega^{-1}=k
\end{eqnarray}
for constant $k$. So there are two cases to consider. If $k=0$ then $\Phi=0$. If $k \neq 0$ then without loss of
generality one can set $\Phi = \Omega$. We shall consider these two cases separately.

\subsection{Solutions with $\Phi \neq 0$}

On setting the conformal factor $\Omega = \Phi$, one finds that the Ricci tensor of the rescaled metric is
\begin{eqnarray}
\label{ricci2a}
{\tilde{R}}_{ij} = \big( (h')^2 +{1 \over 2} \big) \delta_{ij} - {\tilde{\nabla}}_{(i} h'_{j)} - h'_i h'_j
\end{eqnarray}
and moreover ({\ref{cx6}}) can be rewritten as
\begin{eqnarray}
{\tilde{\star}}_3 dh' = - h'
\end{eqnarray}
where ${\tilde{\star}}_3$ denotes the Hodge dual on ${\tilde{{\cal{S}}}}$.
It is then straightforward to show that
\begin{eqnarray}
{\tilde{\nabla}}^2 (h')^2 + (h')^i {\tilde{\nabla}}_i (h')^2 = 2 {\tilde{\nabla}}^{(i} (h')^{j)} {\tilde{\nabla}}_{(i} (h')_{j)} \ .
\end{eqnarray}
Then compactness of ${\tilde{{\cal{S}}}}$ implies that $(h')^2$ is constant, and moreover ${\tilde{\nabla}}_{(i} (h')_{j)}=0$.
So the Ricci tensor of ${\tilde{{\cal{S}}}}$ simplifies to
\begin{eqnarray}
\label{ricci2b}
{\tilde{R}}_{ij} = \big( (h')^2 +{1 \over 2} \big) \delta_{ij}  - h'_i h'_j \ .
\end{eqnarray}
It follows that if $h' \neq 0$, then ${\tilde{{\cal{S}}}}$ is a squashed $S^3$, whereas if $h'=0$, ${\tilde{{\cal{S}}}}$ is a round $S^3$.
Also, note that the Bianchi identity ({\ref{bian}}) can be rewritten as
\begin{eqnarray}
{\tilde{\nabla}}^2 \big( \Phi^{-1} X^I \big) + (h')^i {\tilde{\nabla}}_i \big( \Phi^{-1} X^I \big) =0
\end{eqnarray}
and hence, compactness of ${\tilde{{\cal{S}}}}$ implies that
\begin{eqnarray}
X^I = \Phi Z^I
\end{eqnarray}
for constant $Z^I$.

\subsection{Solutions with $\Phi=0$}
In this case, the Ricci tensor of the conformally rescaled metric is
\begin{eqnarray}
\label{ricci3a}
{\tilde{R}}_{ij} = (h')^2  \delta_{ij} - {\tilde{\nabla}}_{(i} h'_{j)} - h'_i h'_j
\end{eqnarray}
and moreover ({\ref{cx6}}) can be rewritten as
\begin{eqnarray}
dh' = 0 \ .
\end{eqnarray}
Again, one finds that
\begin{eqnarray}
{\tilde{\nabla}}^2 (h')^2 + (h')^i {\tilde{\nabla}}_i (h')^2 = 2 {\tilde{\nabla}}^{(i} (h')^{j)} {\tilde{\nabla}}_{(i} (h')_{j)}
\end{eqnarray}
so compactness of ${\tilde{{\cal{S}}}}$ implies that $(h')^2$ is constant, and moreover ${\tilde{\nabla}}_{(i} (h')_{j)}=0$,
and hence $h'$ is covariantly constant ${\tilde{\nabla}} h' =0$.
So the Ricci tensor of ${\tilde{{\cal{S}}}}$ simplifies to
\begin{eqnarray}
\label{ricci3b}
{\tilde{R}}_{ij} = (h')^2 \delta_{ij}  - h'_i h'_j \ .
\end{eqnarray}
It follows that if $h' \neq 0$, then ${\tilde{{\cal{S}}}}$ is $S^1 \times S^2$, whereas if $h'=0$, ${\tilde{{\cal{S}}}}$ is $T^3$.
Also, note that the Bianchi identity ({\ref{bian}}) can be rewritten as
\begin{eqnarray}
{\tilde{\nabla}}^2 \big( \Omega^{-1} X^I \big) + (h')^i {\tilde{\nabla}}_i \big( \Omega^{-1} X^I \big) =0
\end{eqnarray}
and hence, compactness of ${\tilde{{\cal{S}}}}$ implies that
\begin{eqnarray}
X^I = \Omega Z^I
\end{eqnarray}
for constant $Z^I$.

\newsection{Analysis of Field Equations}

To proceed, we analyse the auxiliary $D$-field equation, which is
\begin{eqnarray}
\label{df}
{1 \over 6} C_{IJK} X^I X^J X^K -1 = -{1 \over 72} c_{2I} \big( {3 \over 4} H_{\mu \nu} F^{I \mu \nu} + D X^I \big) \ .
\end{eqnarray}
Again, we treat the cases $\Phi \neq 0$ and $\Phi=0$ separately. In all cases, it is possible to check directly, using
a  computer calculation, that
the conditions obtained in the previous sections from the analysis of the Killing spinor equations, together
with the $D$-field equation ({\ref{df}}) are sufficient to imply that the Einstein, scalar, gauge, and auxiliary 2-form
equations are satisfied{\footnote{Due to the length of these field equations, we do not list them here; however
they can be found in the Appendix of \cite{systemx1}.}}. It is therefore sufficient to consider the conditions imposed on the solution by ({\ref{df}}).

\subsection{Solutions with $\Phi \neq 0$}

After some manipulation, one can rewrite ({\ref{df}}) as
\begin{eqnarray}
\Phi^3 \bigg( {1 \over 6} C_{IJK} Z^I Z^J Z^K +{1 \over 48} c_{2I} Z^I \big( (h')^2 -1 \big) \bigg) -1
= -{1 \over 72} c_{2I} Z^I \bigg( -{3 \over 2} \Phi^2 (h')^i {\tilde{\nabla}}_i \Phi
\nonumber \\
 +3 \Phi^2 {\tilde{\nabla}}^2 \Phi
-3 \Phi {\tilde{\nabla}}^i \Phi {\tilde{\nabla}}_i \Phi \bigg) \ .
\end{eqnarray}
Observe that if $c_{2I} Z^I=0$ then this expression implies that $\Phi$ is constant, and the
conditions on the spacetime geometry are then equivalent to those found by
\cite{reallbh} for the 2-derivative theory. This solution is the maximally supersymmetric
near-horizon BMPV geometry \cite{bmpv}.

Suppose instead that $c_{2I} Z^I \neq 0$.
On setting $\Phi = e^{-{V \over 3}}$, ({\ref{df}}) can be further simplified to
\begin{eqnarray}
\label{df2a}
{\tilde{\nabla}}^2 V - {1 \over 2} (h')^i {\tilde{\nabla}}_i V = a e^V +b
\end{eqnarray}
where
\begin{eqnarray}
a = -{72 \over c_{2I} Z^I}, \qquad b = {12 \over c_{2I} Z^I} C_{MNP} Z^M Z^N Z^P +{3 \over 2} ((h')^2-1) \ .
\end{eqnarray}
This type of equation has been considered in Appendix C. If $a >0, b \geq 0$, or $a<0, b \leq 0$
then it admits no solutions, and if $a>0, b<0$ then $V$ is constant. If $V$ is constant,
then the solution is the maximally supersymmetric near-horizon BMPV geometry.

For the remaining case $a<0, b>0$, one also finds that ({\ref{df2a}}) can be further simplified to
\begin{eqnarray}
(h')^i {\tilde{\nabla}}_i V =0 \qquad {\tilde{\nabla}}^2 V = a e^V +b \ .
\end{eqnarray}

\subsection{Solutions with $\Phi=0$}

After some manipulation, one can rewrite ({\ref{df}}) as
\begin{eqnarray}
\Omega^3 \bigg( {1 \over 6} C_{IJK} Z^I Z^J Z^K +{1 \over 48} c_{2I} Z^I  (h')^2  \bigg) -1
= -{1 \over 72} c_{2I} Z^I \bigg( -{3 \over 2} \Omega^2 (h')^i {\tilde{\nabla}}_i \Omega
\nonumber \\
 +3 \Omega^2 {\tilde{\nabla}}^2 \Omega
-3 \Omega {\tilde{\nabla}}^i \Omega {\tilde{\nabla}}_i \Omega \bigg) \ .
\end{eqnarray}
Again, if $c_{2I} Z^I=0$ then this expression implies that $\Omega$ is constant, and the
conditions on the spacetime geometry are then equivalent to those found by
\cite{reallbh} for the 2-derivative theory. In particular, in this case, the
solution is either $AdS_3 \times S^2$ if $h \neq 0$, or $\bR^{4,1}$ if $h=0$, and
these solutions are maximally supersymmetric.

Suppose instead that $c_{2I} Z^I \neq 0$.
On setting $\Omega = e^{-{V \over 3}}$, ({\ref{df}}) can be further simplified to
\begin{eqnarray}
\label{df2b}
{\tilde{\nabla}}^2 V - {1 \over 2} (h')^i {\tilde{\nabla}}_i V = a e^V +b
\end{eqnarray}
where
\begin{eqnarray}
a = -{72 \over c_{2I} Z^I}, \qquad b = {12 \over c_{2I} Z^I} C_{MNP} Z^M Z^N Z^P +{3 \over 2} (h')^2 \ .
\end{eqnarray}
From the results of Appendix C,  if $a >0, b \geq 0$, or $a<0, b \leq 0$
then ({\ref{df2b}}) admits no solutions. If $a>0, b<0$ then $V$ is constant,
and the solution is $AdS_3 \times S^2$ if $h \neq 0$, and $\bR^{4,1}$ if $h=0$.

For the remaining case $a<0, b>0$, one also finds that ({\ref{df2b}}) can be further simplified to
\begin{eqnarray}
(h')^i {\tilde{\nabla}}_i V =0 \qquad {\tilde{\nabla}}^2 V = a e^V +b \ .
\end{eqnarray}

\newsection{Summary of Solutions}

In this section, we collate our results and summarise the near-horizon geometries. In addition, as we
have obtained the Ricci tensor for ${\tilde{{\cal{S}}}}$ in ({\ref{ricci2b}}) and ({\ref{ricci3b}}),
it is straightforward to introduce local co-ordinates on ${\tilde{{\cal{S}}}}$ in order to write the
solutions explicitly. The details for this calculation can be found in \cite{reallbh}. 

We remark that we have proven that either $c_{2I} X^I$ vanishes identically, or is never zero.
In the former case, the contribution from the higher derivative terms
vanishes, and the solutions reduce to the maximally supersymmetric near-horizon geometries found in \cite{reallbh, gutbh}.
Hence, for the remainder of this section we shall assume that $c_{2I} X^I \neq 0$.

\subsection{Timelike Solutions with Event Horizon Topology $S^3$}

If $\Delta \neq 0$ then the spatial cross sections of the event horizon are
conformal to a squashed, or round, $S^3$, with metric
\begin{eqnarray}
ds^2_{{{\cal{S}}}} =  e^{{2V \over 3}} \bigg( \lambda \big( (\sigma^1)^2 + (\sigma^2)^2) + \lambda^2 \big( \sigma^3 \big)^2 \bigg)
\end{eqnarray}
where $0 < \lambda \leq 1$ is constant{\footnote{Solutions with $\lambda>1$ might be expected to correspond
to the higher derivative generalisation of over-rotating BMPV black holes. However, as such
solutions do not have regular horizons, these do not appear in our classification.}}, $V$ is a function on ${\cal{S}}$, and
\begin{eqnarray}
\sigma^1 &=& \sin \phi d \theta - \cos \phi \sin \theta d \psi
\nonumber \\
\sigma^2 &=& \cos \phi d \theta + \sin \phi \sin \theta d \psi
\nonumber \\
\sigma^3 &=& d \phi + \cos \theta d \psi
\end{eqnarray}
are left-invariant 1-forms on $SU(2)$ satisfying
\begin{eqnarray}
d \sigma^i = -{1 \over 2} \epsilon^{ijk} \sigma^j \wedge \sigma^k \ .
\end{eqnarray}
The metric on ${\tilde{{\cal{S}}}}$ is 
\begin{eqnarray}
ds^2_{\tilde{{\cal{S}}}} = \lambda \big( (\sigma^1)^2 + (\sigma^2)^2) + \lambda^2 \big( \sigma^3 \big)^2
\end{eqnarray}
with volume form
\begin{eqnarray}
{\tilde{\epsilon}}^{(3)}= \lambda^2 \sigma^1 \wedge \sigma^2 \wedge \sigma^3 \ .
\end{eqnarray}
It is also convenient to define a new radial co-ordinate $\rho$ as
\begin{eqnarray}
\rho = e^{{V \over 3}} r \ .
\end{eqnarray}
With these conventions, the five-dimensional near horizon geometry is
\begin{eqnarray}
ds^2= 2 e^{-{V \over 3}} du  \big( d \rho \pm \sqrt{\lambda-\lambda^2} \rho \sigma^3 -{1 \over 2} \rho^2 e^{-V} du \big)
+ e^{{2V \over 3}} \bigg( \lambda \big( (\sigma^1)^2 + (\sigma^2)^2) + \lambda^2 \big( \sigma^3 \big)^2 \bigg)
\end{eqnarray}
and the scalars $X^I$ and 2-form gauge field strengths $F^I$ are given by 
\begin{eqnarray}
X^I &=& e^{-{V \over 3}} Z^I
\nonumber \\
F^I &=& Z^I \bigg( d (e^{-V} \rho du) \pm \sqrt{\lambda- \lambda^2} \sigma^1 \wedge \sigma^2 \bigg)
\end{eqnarray}
for constants $Z^I$. The auxiliary 2-form $H$ is
\begin{eqnarray}
H =  -d \bigg( e^{-{2V \over 3}} \rho^2 du \bigg) - e^{V \over 3} {\tilde{\star}}_3 \bigg(
\pm \sqrt{\lambda-\lambda^2}  \sigma^3 +{1 \over 3} dV \bigg)
\end{eqnarray} 
and the auxiliary scalar is
\begin{eqnarray}
D = 3 e^{-{2V \over 3}} \bigg( \lambda^{-1} -2 -{1 \over 3} {\tilde{\nabla}}^2 V \bigg) \ .
\end{eqnarray} 

The function $V$ satisfies
\begin{eqnarray}
{\tilde{\nabla}}^2 V = a e^V + b
\end{eqnarray}
where ${\tilde{\nabla}}^2 = {\tilde{\nabla}}^i {\tilde{\nabla}}_i$ is the Laplacian on ${\tilde{{\cal{S}}}}$, and
\begin{eqnarray}
a = -{72 \over c_{2I} Z^I}, \qquad b = {12 \over c_{2I} Z^I} C_{MNP} Z^M Z^N Z^P +{3 \over 2} (\lambda^{-1} - 2)
\end{eqnarray}
are constants. If $\lambda \neq 1$, then as a consequence of the compactness arguments presented in  Appendix C, $V$ 
is a function on $S^2$, i.e. is independent of $\phi$, whereas if $\lambda =1$ then $V$ is a function on the
(round) $S^3$.

If $a >0, b \geq 0$, or $a<0, b \leq 0$
then there are no regular horizons. If $a>0, b<0$ then $V$ is constant and 
the solution is the  maximally supersymmetric near-horizon (higher-derivative) BMPV geometry found in \cite{katmadas}.
Observe that if $a<0$, then by choosing a sufficiently small value of $\lambda$, one can obtain a positive value for $b$.
The status of such solutions remains to be determined.

\subsection{Null Solutions}

 The null solutions, which have $\Delta=0$, split into two sub-cases, according to whether $h' \neq 0$ or $h'=0$
 corresponding to event horizon cross-sections with topology $S^1 \times S^2$ and $T^3$ respectively.

\subsubsection{Null Solutions with Event Horizon Topology $S^1 \times S^2$}

For these solutions, the spatial cross-sections of the horizon are conformal to $S^1 \times S^2$. One can introduce local co-ordinates on ${\cal{S}}$, $\{ \phi, \theta, \psi \}$  such that
\begin{eqnarray}
ds^2_{{{\cal{S}}}} = \lambda e^{{2V \over 3}} \bigg( d \phi^2 + d \theta^2 + \sin^2 \theta d \psi^2 \bigg)
\end{eqnarray}
where $\lambda$ is a positive constant, and $V$ is a function on $S^2$ (i.e. $V=V(\theta,\psi)$).
The metric on ${\tilde{{\cal{S}}}}$ is
\begin{eqnarray}
ds^2_{\tilde{{\cal{S}}}} = \lambda \bigg( d \phi^2 + d \theta^2 + \sin^2 \theta d \psi^2 \bigg)
\end{eqnarray}
with volume form
\begin{eqnarray}
{\tilde{\epsilon}}^{(3)}= \lambda^{3 \over 2} \sin \theta d \phi \wedge d \theta \wedge d \psi \ .
\end{eqnarray}
Again, it is convenient to define a new radial co-ordinate as
\begin{eqnarray}
\rho = e^{{V \over 3}} r \ .
\end{eqnarray}
With these conventions, the five-dimensional near horizon geometry is
\begin{eqnarray}
ds^2 = 2 e^{-{V \over 3}} du \big(d \rho + \rho d \phi \big) + \lambda e^{{2V \over 3}} \bigg( d \phi^2 + d \theta^2 + \sin^2 \theta d \psi^2 \bigg)
\end{eqnarray}
and the scalars $X^I$ and 2-form gauge field strengths $F^I$ are given by 
\begin{eqnarray}
X^I &=& e^{-{V \over 3}} Z^I
\nonumber \\
F^I &=& \lambda^{1 \over 2} Z^I \sin \theta d \theta \wedge d \psi 
\end{eqnarray}
for constants $Z^I$. The auxiliary 2-form $H$ is
\begin{eqnarray}
H =  - e^{V \over 3} {\tilde{\star}}_3 \bigg(d \phi +{1 \over 3} dV \bigg)
\end{eqnarray} 
and the auxiliary scalar is
\begin{eqnarray}
D = 3 e^{-{2V \over 3}} \bigg( \lambda^{-1}  -{1 \over 3} {\tilde{\nabla}}^2 V \bigg) \ .
\end{eqnarray} 

The function $V$ satisfies
\begin{eqnarray}
{\tilde{\nabla}}^2 V = a e^V + b
\end{eqnarray}
where ${\tilde{\nabla}}^2 = {\tilde{\nabla}}^i {\tilde{\nabla}}_i$ is the Laplacian on $S^2$
equipped with metric
\begin{eqnarray}
ds^2 (S^2) = \lambda \bigg( d \theta^2 + \sin^2 \theta d \psi^2 \bigg)
\end{eqnarray}
and
\begin{eqnarray}
a = -{72 \over c_{2I} Z^I}, \qquad b = {12 \over c_{2I} Z^I} C_{MNP} Z^M Z^N Z^P +{3 \over 2} \lambda^{-1} 
\end{eqnarray}
are constants.

If $a >0, b \geq 0$, or $a<0, b \leq 0$
then there are no regular horizons. If $a>0, b<0$ then $V$ is constant and 
the solution is the maximally supersymmetric (higher-derivative)  $AdS^3 \times S^2$ solution of \cite{katmadas}.
Observe that if $a<0$, then by choosing a sufficiently small value of $\lambda$, one can obtain a positive value for $b$.
The status of such solutions remains to be determined. 

\subsubsection{Null Solutions with Event Horizon Topology $T^3$}

For these solutions the spatial cross-sections of the event horizon are conformal to $T^3$. One can introduce local co-ordinates
on $T^3$, $\{ \phi, \theta, \psi \}$ such that
\begin{eqnarray}
ds^2_{{{\cal{S}}}} =  e^{{2V \over 3}} \bigg( d \phi^2 + d \theta^2 + d \psi^2 \bigg)
\end{eqnarray}
where $V$ is a function on $T^3$.
The metric on ${\tilde{{\cal{S}}}}$ is
\begin{eqnarray}
ds^2_{\tilde{{\cal{S}}}} =  d \phi^2 + d \theta^2 + d \psi^2 
\end{eqnarray}
with volume form
\begin{eqnarray}
{\tilde{\epsilon}}^{(3)}= d \phi \wedge d \theta \wedge d \psi \ .
\end{eqnarray}
Again, it is convenient to define a new radial co-ordinate as
\begin{eqnarray}
\rho = e^{{V \over 3}} r \ .
\end{eqnarray}
With these conventions, the five-dimensional near horizon geometry is
\begin{eqnarray}
ds^2 = 2 e^{-{V \over 3}} du d \rho  + e^{{2V \over 3}} \bigg( d \phi^2 + d \theta^2 + d \psi^2 \bigg)
\end{eqnarray}
and the scalars $X^I$ and 2-form gauge field strengths $F^I$ are given by 
\begin{eqnarray}
X^I &=& e^{-{V \over 3}} Z^I
\nonumber \\
F^I &=& 0
\end{eqnarray}
for constants $Z^I$. The auxiliary 2-form $H$ is
\begin{eqnarray}
H =   {1 \over 3} e^{V \over 3} {\tilde{\star}}_3 dV 
\end{eqnarray} 
and the auxiliary scalar is
\begin{eqnarray}
D =- e^{-{2V \over 3}} {\tilde{\nabla}}^2 V  \ .
\end{eqnarray} 

The function $V$ satisfies
\begin{eqnarray}
{\tilde{\nabla}}^2 V = a e^V + b
\end{eqnarray}
where ${\tilde{\nabla}}^2 = {\tilde{\nabla}}^i {\tilde{\nabla}}_i$ is the Laplacian on $T^3$,
and
\begin{eqnarray}
a = -{72 \over c_{2I} Z^I}, \qquad b = {12 \over c_{2I} Z^I} C_{MNP} Z^M Z^N Z^P 
\end{eqnarray}
are constants. 

If $a >0, b \geq 0$, or $a<0, b \leq 0$
then there are no regular horizons. If $a>0, b<0$ then $V$ is constant and 
the solution is $\bR^{1,4}$.
The status of solutions with $a<0, b>0$ remains to be determined.

 \newsection{Conclusions}

In this paper, we have classified all supersymmetric extremal near-horizon geometries
of supersymmetric black hole solutions in the higher derivative $N=2, D=5$ 
supergravity constructed in \cite{hanaki}. We have proven that either $c_{2I} X^I$ vanishes
identically on the horizon, or it never vanishes. In the former case, the near-horizon solutions
are the maximally supersymmetric near-horizon solutions already known in the two-derivative theory
\cite{reallbh, gutbh}.
In the latter case, we have found all possible supersymmetric near-horizon solutions,
and we have shown that the spatial cross-sections of the event horizon are conformal to either a squashed or round
$S^3$, $S^1 \times S^2$, or $T^3$. In all cases, the conformal factor is determined in terms of a function
$V$ satisfying a non-linear PDE of the form
\begin{eqnarray}
\label{tb}
{\tilde{\nabla}}^2 V = a e^V + b
\end{eqnarray}
where $a$, $b$ are constants determined in terms of the near-horizon data, as described in the previous section,
with $a \neq 0$. The sign of $a$ is identical to that of $c_{2I} X^I$. 
The function $V$ is either defined on the (round) $S^3$, or $S^2$, or $T^3$, and ${\tilde{\nabla}}^2$ is the appropriate
Laplacian in each case. Equation ({\ref{tb}}) has been examined in \cite{taubes, kazdan, bradlow, manton}.

If $a >0, b \geq 0$, or $a<0, b \leq 0$ then there is no solution to the equation ({\ref{tb}}),
and there are no regular horizons. If $a>0, b<0$ then $V$ is constant and 
the solution  reduces to one of the maximally supersymmetric solutions
found in \cite{katmadas}.

However, the most interesting case arises when $a<0, b>0$.
In particular, for the solutions with event horizon topology $S^3$ or $S^1 \times S^2$, we have
shown that provided $c_{2I} X^I$ is negative, one can always arrange for $b>0$ by choosing another parameter
(corresponding to the angular momentum associated with $h'$) to be sufficiently small.
It is clear that if $a<0, b>0$ then   ({\ref{tb}}) admits a solution for which $V$ is constant, and for which
the geometry is again one of the maximally supersymmetric near-horizon solutions of \cite{katmadas}.
It is however far from clear that this solution to ({\ref{tb}}) is unique, when $a$ and $b$ have this choice of sign. 
Notwithstanding this,
we are also not aware of any explicit globally well-defined and regular non-constant solutions
to ({\ref{tb}}) when $a<0, b>0$. If such solutions were to exist, then they would describe supersymmetric 
black holes with scalar hair on the horizon. They would also lie outside of the classification of
solutions given in \cite{extr1}, because solutions with non-constant $V$ would not have horizons with two
commuting rotational isometries. 

Another interesting issue is the possible extension of the analysis to construct a uniqueness
theorem for supersymmetric black holes. In the two-derivative theory, it has been shown
in \cite{reallbh, gutbh} that the only supersymmetric black holes with event horizon topology
$S^3$ are the BMPV black holes. We remark that no corresponding uniqueness theorem exists
for black rings with event horizon topology $S^1 \times S^2$. Although no analytic solution is currently known
for black holes with event horizon topology $S^3$ in the higher derivative theory, it is reasonable to expect that such solutions exist.
One can use supersymmetry to constrain the bulk geometries of these solutions.

The uniqueness proof for the 2-derivative theory considers
the case for which the Killing vector ${\partial \over \partial u}$ 
constructed from the Killing spinor is timelike, both in the near-horizon limit solution, and in the bulk geometry. 
The analysis proceeds by recalling that one can write 
the 5-dimensional solution as a $U(1)$ fibration over
a 4-dimensional hyper-K\"ahler base space $HK$
\begin{eqnarray}
\label{fibr}
ds^2 = -f^2 (du + \omega)^2 + f^{-1} ds^2_{HK}
\end{eqnarray}
where $f$ is a $u$-independent function, $ds^2_{HK}$ is a $u$-independent hyper-K\"ahler metric on the base
space, and $\omega$ is a $u$-independent 1-form on $HK$. First, it was shown that for the near-horizon geometry
of solutions with $S^3$ horizon topology,
$HK$ is $\bR^4$. Given this, together with sufficient assumptions of regularity outside the event horizon, it was
then shown that the base space for the bulk black hole solution must be $\bR^4$.
Following on from this, one can also write the
gauge field strengths as
\begin{eqnarray}
\label{sdg}
F^I = -d \big(f^2 X^I (du + \omega) \big) + \Theta^I
\end{eqnarray}
where $\Theta^I$ are harmonic 2-forms on $HK$, and for the near-horizon geometry one finds $\Theta^I=0$.
This  can be extended into the bulk black hole solution, for which one must have $\Theta^I=0$.

It is straightforward to show that exactly the same results hold for the higher derivative theory considered here.
In particular, as a consequence of the classification of the timelike supersymmetric solutions in \cite{larsen},
it is known that the metric can be written as a $U(1)$ fibration over a 4-dimensional hyper-K\"ahler base space $HK$
as in ({\ref{fibr}}). For the near-horizon solutions constructed here, one finds that
\begin{eqnarray}
f= e^{-{2 V \over 3}} \rho , \qquad \omega = - \rho^{-2} e^V \bigg( d \rho \pm \sqrt{\lambda-\lambda^2} \rho (d \phi+\cos \theta d \psi) \bigg)
\end{eqnarray}
and
\begin{eqnarray}
ds^2_{HK} &=& {1 \over \rho}  \bigg( d \rho \pm \sqrt{\lambda-\lambda^2} \rho (d \phi+\cos \theta d \psi) \bigg)^2
\nonumber \\
&+&\rho \bigg(\lambda^2  (d \phi+\cos \theta d \psi)^2 + \lambda (d \theta^2 + \sin^2 \theta d \psi^2) \bigg) \ .
\end{eqnarray}
This metric  is flat, and so just as in the two derivative theory, one finds that the hyper-K\"ahler base space
for the full black hole solution must be $\bR^4$. 
Furthermore, one can also write the gauge field strengths for the timelike solutions as in ({\ref{sdg}}), where $\Theta^I$ are
harmonic self-dual 2-forms on $HK$. For the near-horizon geometry constructed here,
one finds that $\Theta^I=0$, and hence using exactly the same reasoning as in the analysis for the 2-derivative theory, one finds
that $\Theta^I=0$ for the full black hole solution.

The requirement that $HK=\bR^4$ and $\Theta^I=0$ imposes significant constraints on any possible black hole solutions.
It remains to analyse the remaining field equations, which in spite of the simplification described above, remain
somewhat non-trivial. The extent to which this analysis will produce a uniqueness theorem will depend on the existence
(or non-existence) of non-constant solutions to  ({\ref{tb}}) when $a<0, b>0$. Work on this is in progress.

 \setcounter{section}{0}

\appendix{Spin Connection}

The non-vanishing components of the spin connection associated with the basis ({\ref{basis}}) are
\begin{eqnarray}
\omega_{+,+-}&=& -r \Delta, \quad \omega_{+,+i}= {r^2 \over 2} \big(\Delta h_i
- \partial_i \Delta \big), \quad \omega_{+,-i}=-{1 \over 2} h_i,
\quad \omega_{+,ij}=-{r \over 2} (dh)_{ij}
\nonumber \\
\omega_{-,+i}&=& -{1 \over 2} h_i
\nonumber \\
\omega_{i,+-}&=& {1 \over 2} h_i, \quad \omega_{i,+j}= -{r \over 2} (dh)_{ij}, \quad
\omega_{i,jk}={\tilde{\omega}}_{i,jk}
\end{eqnarray}
where ${\tilde{\omega}}_{i,jk}$ is the spin connection of ${\cal{S}}$.

\appendix{Spinorial Geometry Conventions}

Spinorial geometry techniques were originally developed to analyse supersymmetric solutions
of ten and eleven dimensional supergravity \cite{tend, elevend}. Here we apply them to five-dimensional
supergravity.
The space of Dirac spinors consists of the space of complexified forms on $\bR^2$,
which has basis $\{ 1, e_1, e_2, e_{12}=e_1 \wedge e_2 \}$.
We define the action of the Clifford algebra generators on this space via
\begin{eqnarray}
\gamma_i = -e_i \wedge  - i_{e_i}, \qquad \gamma_{i+2}= i \big(-e_i \wedge + i_{e_i} \big) \quad i=1,2
\end{eqnarray}
and set
\begin{eqnarray}
\gamma_0 = i \gamma_{1234}
\end{eqnarray}
which acts as
\begin{eqnarray}
\gamma_0 1 = i 1 , \quad \gamma_0 e_{12} = i e_{12}, \quad \gamma_0 e_i = -i e_i \ .
\end{eqnarray}
We then define generators adapted to the frame ({\ref{frame}}) as
\begin{eqnarray}
\Gamma_\pm = {1 \over \sqrt{2}} (\gamma_3 \pm \gamma_0),
\quad \Gamma_1 = \gamma_1, \quad \Gamma_2 = \sqrt{2} e_2 \wedge,
\qquad \Gamma_{{\bar{2}}} = \sqrt{2} i_{e_2}
\end{eqnarray}
where we take a basis $\{ {{\bf{e}}}^1, {{\bf{e}}}^2, {{\bf{e}}}^{{\bar{2}}} \}$ for ${\cal{S}}$ such that ${{\bf{e}}}^{{\bar{2}}}= ({{\bf{e}}}^2)^*$ and
\begin{eqnarray}
ds^2_{{\cal{S}}} = ({{\bf{e}}}^1)^2 +2 {{\bf{e}}}^2 {{\bf{e}}}^{{\bar{2}}} \ .
\end{eqnarray}
With these conventions, the space of positive chirality spinors is spanned by
$\{ 1-e_1, e_2+e_{12} \}$, and the space of negative chirality spinors is spanned by
$\{ 1+e_1, e_2-e_{12} \}$ and we remark that $Spin(3)$, with generators
$i \Gamma_{2 {{\bar{2}}}}, \Gamma_1(\Gamma_2+\Gamma_{{\bar{2}}}),
i  \Gamma_1(\Gamma_2-\Gamma_{{\bar{2}}})$ form a representation of $SU(2)$ acting on
$\{1-e_1, e_2+e_{12} \}$.

A $Spin(4,1)$ invariant inner product $B$ on the space of spinors is then given by
\begin{eqnarray}
\label{prod}
B (\epsilon_1, \epsilon_2) = \langle \gamma_0 \epsilon_1, \epsilon_2 \rangle
\end{eqnarray}
where $\langle , \rangle$ denotes the canonical inner product on $\bC^4$ equipped with basis
$\{1, e_1, e_2, e_{12} \}$.

\appendix{A Vortex Equation}

Suppose $M$ is a compact manifold without boundary, $\kappa$ is an isometry of $M$,
and $V$ is a smooth function on $M$ satisfying
\begin{eqnarray}
\label{liou}
\nabla^2 V  -{1 \over 2} \kappa^i \nabla_i V = a e^V +b
\end{eqnarray}
for constants $a, b$, $a \neq 0$, where $\nabla$ is the Levi-Civita connection, and
$\nabla^2 V = \nabla_i \nabla^i V$.

First consider the cases for which $a >0$ and $b \geq 0$, or $a <0$ and $b \leq 0$.
Note that
\begin{eqnarray}
\label{van1}
\int_M  a e^V + b = \int_M \nabla^2 V -{1 \over 2} \kappa^i \nabla_i V =0
\end{eqnarray}
on integrating by parts. However, as $ae^V+b$ is either everywhere positive or
negative, this leads to a contradiction. Hence there are no solutions in these two cases.

Suppose instead that $a>0$ and $b<0$. Note that $V$ attains a global minimum at $p \in M$.
At $p$, $\kappa^i \nabla_i V =0$, and $\nabla^2 V = \alpha \geq 0$.
It follows that at $p$,
\begin{eqnarray}
e^V = {\alpha - b \over a}
\end{eqnarray}
and hence
\begin{eqnarray}
e^V \geq  {\alpha - b \over a}
\end{eqnarray}
everywhere on $M$. In particular, one then finds $a e^V +b \geq 0$ everywhere on $M$.
It then follows as a consequence of ({\ref{van1}}) that $ae^V +b=0$, i.e. $V$ is constant.

Next suppose that $a<0$ and $b>0$.
We shall consider two cases which are of particular importance in the context
of the black hole solutions, and in both cases we take $M$ to be 3-dimensional.

In the first case, suppose that
\begin{eqnarray}
\label{ex1}
R_{ij} = \big(\kappa^2+{1 \over 2} \big) \delta_{ij} - \kappa_i \kappa_j , \qquad \nabla_i \kappa_j = -{1 \over 2}
\epsilon_{ij}{}^k \kappa_k \ .
\end{eqnarray}
Note that this expression for the Ricci tensor implies that
\begin{eqnarray}
\nabla_i \big(\nabla^2 V \big) = - (\kappa^2+{1 \over 2}) \nabla_i V + \kappa_i \kappa^\ell \nabla_\ell V
+ \nabla^2 \nabla_i V \ .
\end{eqnarray}
It follows that
\begin{eqnarray}
\label{cb1}
\int_M \kappa^i \nabla_i V \nabla^2 V &=& - \int_M V \kappa^i \nabla_i (\nabla^2 V)
\nonumber \\
&=& - \int_M V \kappa^i \big(- (\kappa^2+{1 \over 2}) \nabla_i V + \kappa_i \kappa^\ell \nabla_\ell V
+ \nabla^2 \nabla_i V \big)
\nonumber \\
&=& -\int_M V \kappa^i \nabla^2 \nabla_i V
\end{eqnarray}
where we have made use of the fact that ({\ref{ex1}}) implies that $\kappa^2$ is constant.
However, note also that
\begin{eqnarray}
\label{cb2}
\int_M \kappa^i \nabla_i V \nabla^2 V &=& \int_M V \nabla^2 \big( \kappa^i \nabla_i V \big)
\nonumber \\
&=& \int_M V \kappa^2 \nabla^2 \nabla_i V + 2 V \nabla^\ell \kappa^i \nabla_\ell \nabla_i V + V \nabla^2 \kappa^i \nabla_i V
\nonumber \\
&=& \int_M V \kappa^i \nabla^2 \nabla_i V
\end{eqnarray}
where again ({\ref{ex1}}) has been used to rewrite $\nabla^2 \kappa_i=-{1 \over 2} \kappa_i$.
On comparing, ({\ref{cb1}}) with ({\ref{cb2}}) one finds
\begin{eqnarray}
\label{cb3}
\int_M \kappa^i \nabla_i V \nabla^2 V =0 \ .
\end{eqnarray}
Next, note that ({\ref{liou}}) implies
\begin{eqnarray}
\int_M \kappa^i \nabla_i V \big(\nabla^2 V -{1 \over 2} \kappa^j \nabla_j V \big)
= \int_M \kappa^i \nabla_i V \big(a e^V + b \big)
\end{eqnarray}
and note that on partially integrating, the contribution from the RHS vanishes,
and also ({\ref{cb3}}) implies that the contribution from the first term on the LHS also vanishes.
Hence
\begin{eqnarray}
\int_M (\kappa^i \nabla_i V)^2 =0
\end{eqnarray}
so
\begin{eqnarray}
\kappa^i \nabla_i V =0 \ .
\end{eqnarray}
Hence, if $\kappa \neq 0$, then one finds that ({\ref{liou}}) simplifies further to
\begin{eqnarray}
\label{liou2}
{{\cal{L}}}_\kappa V =0, \qquad \Box V = ae^V + b
\end{eqnarray}
where $\Box$ denotes the Laplacian on $S^2$.

In the second case, suppose that
\begin{eqnarray}
\label{ex2}
R_{ij} =  \kappa^2\delta_{ij} - \kappa_i \kappa_j , \qquad \nabla \kappa =0 \ .
\end{eqnarray}
Using essentially the same reasoning used to treat the previous case, one again finds
that ({\ref{liou}}) can be simplified to ({\ref{liou2}}).

\section*{Acknowledgments}
The work of W. S. was supported in part by the National Science Foundation under grant number PHY-0903134. J. G. is supported by the EPSRC grant EP/F069774/1. P. S. and D. K. are supported by INFN.
J. G.  would like to thank S. Beheshti, M. Dunajski, G. W. Gibbons, N. Manton and C. Papageorgakis
for useful discussions.


\end{document}